\documentclass{biophys}
\usepackage{helvet,times}
\usepackage{bm,textcomp}
\usepackage{amsmath}
\usepackage{amssymb}
\usepackage{bm}
\usepackage{calc}
\usepackage[round,numbers,sort&compress]{natbib}
\usepackage[ansinew]{inputenc}

\usepackage{color}
\definecolor{darkred}{rgb}{0.5,0,0}
\definecolor{darkgreen}{rgb}{0,0.5,0}
\definecolor{darkblue}{rgb}{0,0,0.5}

\newcommand{\old}[1]{}

\renewcommand{\emph}[1]{{\it #1}}

\jno{kxl014}
\gridframe{N}
\cropmark{Y}
\doi{doi: 10.1529/biophysj.106.089789}

\begin{document}
\title{The robust assembly of small symmetric nano-shells}
\author{Jef Wagner$^\ast$, Roya Zandi$^\ast$}
\address{$^\ast$Department of Physics and Astronomy, University of
California, Riverside, 900 University Ave, Riverside, California
92521, USA}

\begin{abstract}
{Highly symmetric nano-shells are found in many biological systems,
such as clathrin cages and viral shells. Several studies have shown that
symmetric shells appear in nature as a result of the free energy
minimization of a generic interaction between their constituent
subunits. We examine the physical basis for the formation of
symmetric shells, and using a minimal model we demonstrate that these
structures can readily grow from identical subunits under non
equilibrium conditions. Our model of nano-shell assembly shows that
the spontaneous curvature regulates the size of the shell while the
mechanical properties of the subunit determines the symmetry of the
assembled structure. Understanding the minimum requirements for the
formation of closed nano-shells is a necessary step towards
engineering of nano-containers, which will have far reaching impact
in both material science and medicine.}
{Submitted July 16th, 2014.}
{*Correspondence: }
\end{abstract}

\maketitle
\markboth{Wagner and Zandi}{Robust Assemply of Symmetric Nano-Shells}


\section{INTRODUCTION}
\begin{figure}
  \includegraphics[width=8.7cm]{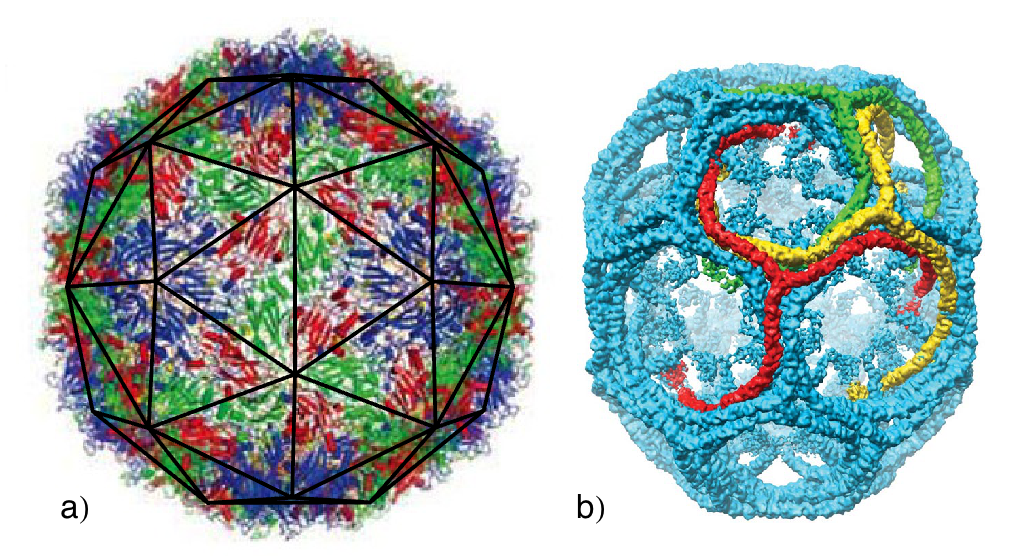}
  \caption{\label{fig:RealShells} (color online) Images of
    (a) a viral shell (enterovirus 71) and (b) a small clathrin coat.
    The viral shell is a T=3 icosahedral virus, constructed of 180
    identical protein subunits.  The proteins sit in three color
    coded quasi-equivalent positions, grouped into triangles as seen
    with the dark outline in subfigure (a). The clathrin coat (b) has
    a $D_6$ dihedral symmetry and is often referred to as a $D_6$
    barrel. The clathrin molecules have a triskelion (three legged)
    shape, that bind anti-parallel leg-to-leg. The symmetries
    displayed by these shells allow for very detailed imaging through
    the technique of spatial averaging.}
\end{figure}

The spontaneous assembly of building blocks into complex ordered
structures is ubiquitous in nature. Examples include the formation of
symmetric shells with extraordinary accuracy such as clathrin
vesicles for membrane trafficking in cells and viral shells for the
protection and transportation of viral genome. Based on very basic
physical principles, it has been argued that small spherical viruses
display structures with icosahedral symmetry\cite{Crick:1952a}, and
while some larger clathrin vesicles are irregular in shape, the
smallest clathrin shells observed {\it in vitro} tend to
take on a discrete set of symmetric structures\cite{Crowther:1976a}.
Figure~\ref{fig:RealShells} shows the 3D image reconstructions of
enterovirus 71, an icosahedral virus\cite{Hogle:2012a} and a $D_6$
symmetric clathrin coat\cite{Fotin:2004a} obtained through the use of
spatial averaging over the set of associated symmetries.

The simplicity and highly symmetric shape of viral shells with
diameters ranging from 20 to 120 $nm$ have in particular attracted
the attention of scientists for many decades\cite{Caspar:1962a,
Hicks:2006a, Zandi:2004a, Bruinsma:03a, Lidmar:03a, Hagan:06a,
Chen:2007a}.  The simplest viruses are made of a genome encapsulated
in a protein shell called the capsid. Quite remarkably, under many
circumstances, the viral coat proteins assemble spontaneously around
its genome or other negatively charged cargoes to form capsids
identical to those observed {\it in vivo}\cite{Cadena:2012a,
Ni:2012a, Lin:2012a, Zlotnick:1994a, Siber:2010a}.  Despite the
importance of engineered biological nano-shells in gene and drug
delivery and other biomedical technologies, the mechanisms and
factors that control the structure and stability of shells made of
identical building blocks are just beginning to be understood.
Constructing the building blocks with interactions similar to those
in the natural nano-containers may lead to the fabrication and design
of precise synthetic nano-structures
\cite{Chen:2007a, Manoharan:2003a, Li:2005a}.

There have been many studies investigating the equilibrium shapes of
shells formed from one or two identical subunits under external
constraints\cite{ Zandi:2004a,Bruinsma:03a,Lidmar:03a,Chen:2007a,
Siber:2010a,Twarock:2008a,Luque:12a, Vernizzi:2007a, Vernizzi:2011a}.
The simple case of spherical colloids or circular disks constrained
to reside on the surface of a sphere shows that the stability of
formed shells depends strongly on the number of assembly subunits and
interactions between them. For example, the solutions of the optimal
packing problem of N identical hard disks on the surface of a sphere,
known as the Tammes problem, reveal the presence of a number of local
maxima in the plot of the sphere coverage vs. number of disks, $N$.
The structures characterized by ``Magic Numbers'', $N=12, 24, 32,
44,\ldots$ corresponding to the local maximum coverage often have
higher symmetry than their neighboring structures\cite{Bruinsma:03a}.
A different set of magic numbers appear when minimizing the free
energy of the $N$ identical disks interacting through Lennard Jones
potential. Some of these magic numbers and their associated shells
coincide with the number of capsomers (protein multimers) in
structures displaying icosahedral symmetry. Magic numbers
corresponding to non-icosahedral shells with octahedral and cubic
symmetries also appear in Monte Carlo simulations of disks sitting
on the surface of a sphere\cite{Bruinsma:03a, Zandi:2004a},
consistent with the structures observed in virus assembly
experiments preformed {\it in vitro}.

The same magic numbers and structures were observed in the Monte
Carlo simulations of the self-assembly of cone-shaped particles with
attractive interactions\cite{Chen:2007a}. The authors in
Ref.~\cite{Chen:2007a} showed that over a range of cone angles, a
unique precise sequence of robust clusters form under equilibrium
conditions. Note that the self-assembly of attractive spheres under a
spherical convexity constraint reproduces exactly the same sequence
of shells\cite{Chen:2007b}. The structure and symmetry of the small
cluster sizes in these simulations\cite{Chen:2007a,Chen:2007b} were
identical to those observed in the experimental studies of
evaporation-driven assembly of colloidal
spheres\cite{Manoharan:2003a}.

Quite interestingly, some of these magic numbers also appeared in
completely different experiments, which showed that under appropriate
conditions clathrin molecules spontaneously assemble {\it in vitro}
to form highly symmetric vesicles whose radii depend on the size of
cargoes\cite{Crowther:1976a}. The fact that in distinct systems we
observe shells with identical symmetry reveals the existence of a
common underlying design principle governing the assembly of these
structures.

\begin{figure*}
  \includegraphics[width=17.8cm]{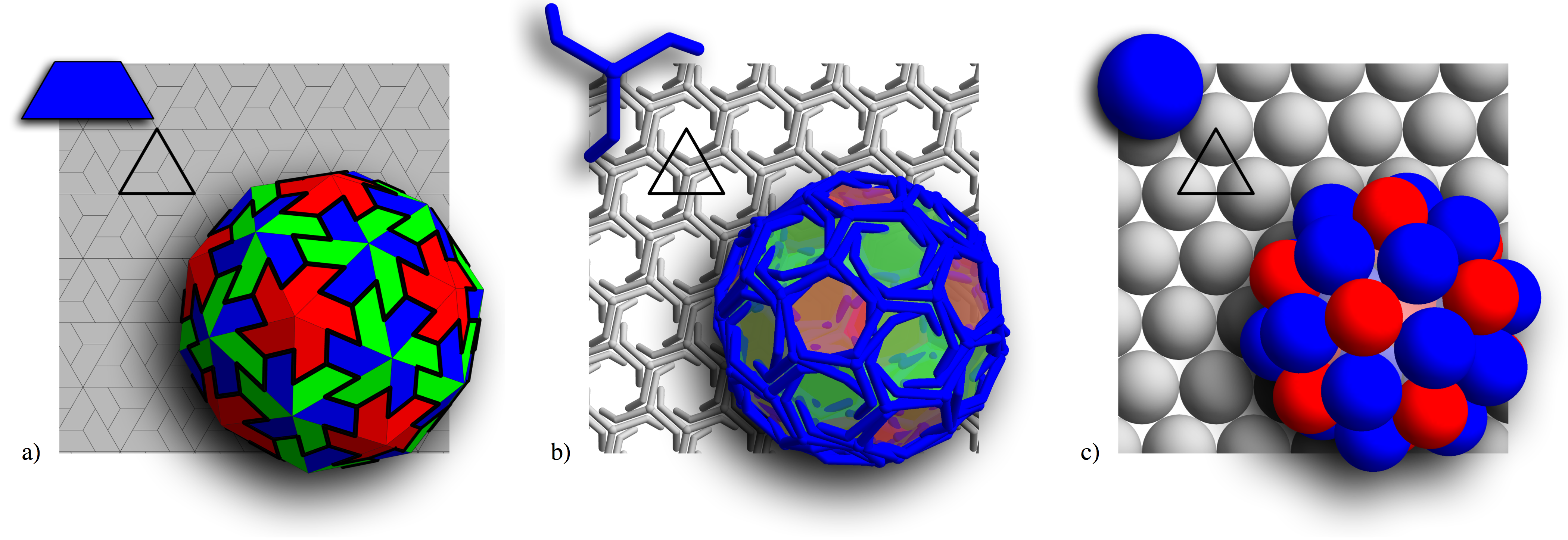}
  \caption{\label{fig:ShellTypes} (color online) Three different
    systems that display a shell with the exact same symmetry. The
    black triangle in the background of each subfigure highlights the
    triangular subunits. Shell (a) is constructed of trimers similar
    to the viral shells, shell (b) is a soccer ball clathrin coat
    constructed from triskelion molecules, and shell (c) is built of
    small spheres connected through Lennard Jones potential with a
    convexity constraint. As seen in the background, all the subunits
    display a local hexagonal symmetry if packed onto a flat
    surface.}
\end{figure*}

One can readily observe that the building blocks of all the
aforementioned systems can be packed with hexagonal symmetry in a
flat space. This is clearly illustrated in Fig.~\ref{fig:ShellTypes},
which shows a viral capsid, a clathrin vesicle, and a shell
constructed of spheres, all with their associated building blocks.
All shells in Fig.~\ref{fig:ShellTypes} have icosahedral symmetry.
While defect free stacks of hexagonal layers in the 3 dimensional
space can easily grow from interacting Lennard Jones particles under
non-equilibrium conditions, a closed shell requires the formation of
12 defects with local fivefold rotational symmetry (pentamers). In
symmetric shells the position of pentamers with respect to each other
is precise, {\it i.e.}, the symmetry can be readily broken if one
pentamer is slightly misplaced. One would then expect that the
assembly of highly symmetric shells proceeds reversibly. In fact,
most previous studies and simulations were done assuming the process
of assembly is completely reversible, the switching between pentamers
and hexamers can easily take place when necessary\cite{Hagan:06a,
Zlotnick:1994a, Schwartz:1998a, Schwartz:2000a, Rapaport:04a,
Nguyen:07a}.

In this paper, we investigate the growth of a shell by subunits that
can connect together with a local hexagonal symmetry. Despite the
sensitivity of the symmetry of shells to the exact location of
pentamers, under many experimental conditions the self-assembly of
perfect icosahedral virus capsids is robust and efficient. The
robustness of the process raises the question of whether a fully
reversible assembly pathway is necessary for the formation of
symmetric nano-containers. To this end, we develope a simple model to
investigate the assembly of shells constructed from identical
subunits with the minimum set of designing principles for
irreversible growth.  Unexpectedly, we find that the shells under
non-equilibrium conditions grow to form the highly symmetric
structures (see Figs.~\ref{fig:ClathrinShells},
\ref{fig:VirusShells}, and \ref{fig:SphereShells} below) observed in
biological systems and in equilibrium studies\cite{Zandi:2004a,
Chen:2007a, Chen:2007b}.
  
It is important to note that while there is a striking similarity
between the aforementioned disparate systems, there are also
significant differences. Viral capsids {\it in vivo} tend to only
form structures with icosahedral symmetry, while the clathrin
molecules as well as conical particles form shells with additional
symmetries. The goal of the paper is to explain the basis of the 
similarities and differences displayed in nature's nano-containers.
According to our studies, the differences in these systems are the
result of the distinct mechanical properties of their constituant
building blocks. Our findings elucidate the kinetic pathways of
assembly and fundamental packing principles in curved space observed
in the distinct biological structures.

\section*{MODELS AND METHODS}
To study the kinetic pathways of the shell growth, we employed
triangular subunits, which can represent either a trimer of viral
capsid proteins, a triskelion molecule, or locus of three disks, see
the highlighted triangles in Fig.~\ref{fig:ShellTypes}. In the model,
the growth proceeds through the irreversible addition of the
triangular subunits to the exterior edges of the incomplete shell.
After the addition of each subunit, the elastic sheet is allowed to
relax and to find its minimum energy configuration.  The minimization
of the elastic energy of the growing shell is done numerically using
a non-linear conjugate gradient method\cite{NumericalRecipes}.

\begin{figure}[b]
  \begin{center}
    \includegraphics{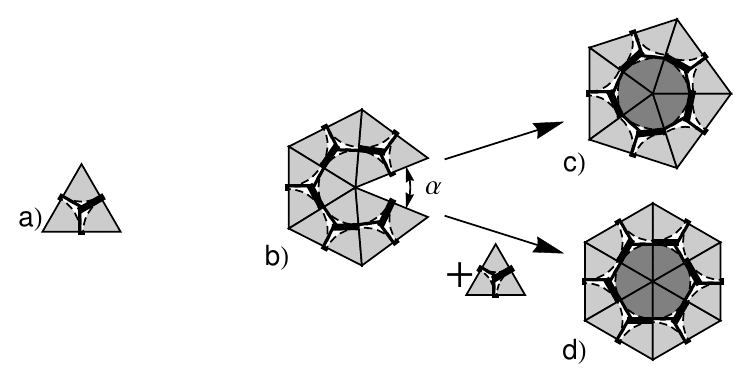}
  \end{center}
  \caption{\label{fig:subunits} A subunit illustrated as an
    equilateral triangle (a). The subunits bond together edge to
    edge, and the growth of the shell proceeds by adding the subunit
    to the location with the smallest opening angle, $\alpha$ (b). As
    the shell grows, the two unbound edges can either bind together
    to form a pentamer (c), or a new subunit can be added and bound
    along both edges to form a hexamer (d). The choice between
    forming a pentamer and a hexamer is based on which leads to a
    lower energy per subunit in the growing shell.}
\end{figure}

The 2-dimensional growing shell is a bond network built from
triangles representing the smallest subunit of the shell, see
Fig.~\ref{fig:subunits}a. While the subunit in
Fig.~\ref{fig:subunits}a corresponds to the locus of three disks, any
other trimers illustrated in Fig.~\ref{fig:ShellTypes} could be used.
The energy of the triangular shell can then be separated into the
bond stretching and bending energies\cite{ Hicks:2006a,Lidmar:03a}.
The stretching energy comes from deforming the subunits from the
preferred shape of an equilateral triangle, and is modeled by
considering each bond as a linear spring with spring constant $k_s$.
The stretching energy is then simply a sum of the deformation energy
over all triangles
\begin{equation}
  E_s = \sum_{i}\sum_{a=1}^3
  \frac{k_s}{2}(b^a_i-b_0)^2.
\end{equation}
where $i$ indexes the triangular subunits, $b_0$ is the
equilibrium length of the bonds, and $b^a_i$ is the length of the
$a^{\text{th}}$ bond in the $i^{\text{th}}$ subunit. The bending
energy results from the deviation of local radius of curvature from
the preferred one and is calculated by summing over all neighboring
pairs of triangular subunits
\begin{equation}
  E_b = \sum_{\langle i j \rangle}k_b
  \big(1-\cos(\theta_{ij}-\theta_0)\big),
\end{equation}
where $\langle i j\rangle$ indexes pairs of joined subunits, $k_b$ is
the torsional spring constant, and $\theta_0$ is the preferred angle
between two subunits determined by the spontaneous radius of
curvature $R_0$. The angle between neighboring subunits,
$\theta_{ij}$, is defined by the relation $\cos \theta_{ij}=\hat{n}_i
\cdot \hat{n}_i$, where $\hat{n}_i$ is the normal vector for the
$i^{\text{th}}$ subunit. The preferred angle $\theta_0$ is defined as
the angle between the normal vectors of two equilateral triangles
sharing a side of length $b_0$ and whose vertices sit on a sphere of
radius $R_0$. The explicit relation between the preferred angle and
radius of curvature is $\sin \theta_0/2 = (12 R_0^2/b_0^2 -
3)^{-1/2}$.

The total energy of the two dimensional bond network depends upon two
dimensionless parameters, $\bar{\gamma}=k_s b_0^2/k_b$ and $R_0/b_0$.
The parameter $\bar{\gamma}$, the Foppl von Karman (FvK) number,
presents the relative difficulty of deforming a subunit from its
equilateral triangular shape vs. the difficulty of bending away from
the preferred radius of curvature. The second dimensionless parameter
$R_0/b_0$ is simply the spontaneous radius of curvature. Note that
the FvK number in this work is normalized with respect to size of the
subunits $b_0$.

The shell is assumed to grow along the minimum free energy path, so
at each growth step a new subunit is added to the growing edge such
that it maximizes the number of neighbors at the vertices of the
newly accreted subunit. This is accomplished in practice by adding
the new subunit to the location with the smallest opening angle
$\alpha$, see Fig.~\ref{fig:subunits}b. If a vertex on
the edge of a growing shell has already 5 triangles attached then the
assembly could proceed in two different ways: (i) joining the two
neighboring edges without adding a new subunit and thus forming a
pentamer as in Fig.~\ref{fig:subunits}c, or (ii) inserting a new
subunit and constructing a hexamer as in Fig.~\ref{fig:subunits}d.
The choice between forming a pentamer or a hexamer is made based on
which configuration induces a lower energy per subunit in the elastic
sheet.

In addition to the deterministic simulations described above, we
introduced stochasticity into the simulated growth. For the
deterministic growth each subunit is added to the location with the
smallest opening angle, $\alpha$. However, for the stochastic model,
each position for addition of next protein is weighted based on the
following criteria: (i) the smallest opening angle is considered the
most probable one, (ii) opening angles within some range ($\alpha +
\sigma$) of the minimum are assumed to be about equally probable, and
(iii)  the opening angles much larger than the minimum should be very
unlikely.  A Gaussian weighting function $w_i$ centered on the
minimum angle, $\alpha$ with width $\sigma$ has all these properties
\begin{equation} \label{eq:stochastic}
  w_i = e^{-\frac{1}{2}
    \frac{(\alpha_i-\alpha)^2}{\sigma^2}},
\end{equation}
with $\alpha_i$ the opening angle at the $i^{\text{th}}$ vertex along
the exterior edge. The location of the new subunit is then randomly
chosen from the weighted list of possible locations.  In the
stochastic model we still let the shell relax and reach its minimum
elastic energy configuration after the addition of each subunit, and
choose between pentamers and hexamers in the same manner as in the
deterministic model. It can be noted at this point that the
deterministic model is simply the stochastic model in the limit as
$\sigma$ approaches zero. For nonzero $\sigma$ the stochastic nature
of this second model allows for the possibility of multiple different
shells to be grown from the same set of parameters.

\section*{RESULTS AND DISCUSSION}
\begin{figure}
  \includegraphics{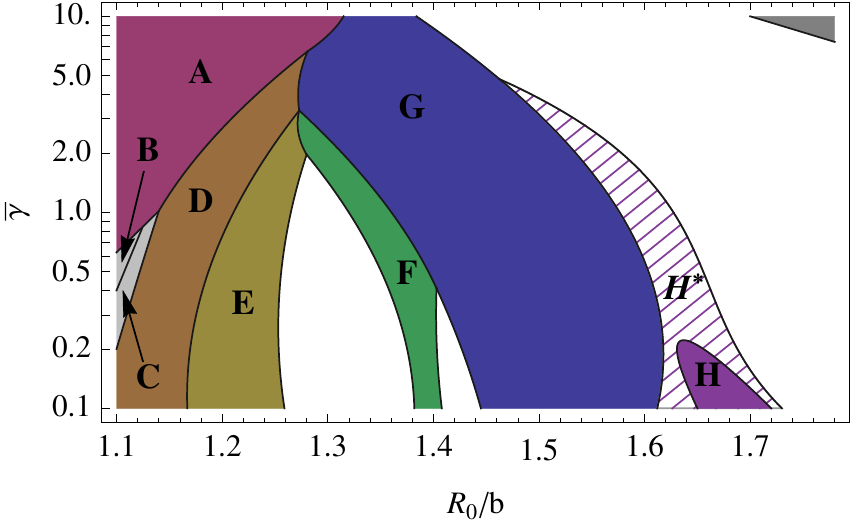}
  \caption{\label{fig:paramspace} (color online) A parameter space
    plot for the deterministic model showing the different kind of
    shells grown for values of the dimensionless parameters
    $\bar{\gamma}$ and $R_0/b_0$. The dark shaded contiguous regions
    labeled by letters (A) through (H) correspond to regions where
    only a single type of symmetric shell is grown. These structures
    are shown in Figs.~\ref{fig:ClathrinShells},
    \ref{fig:VirusShells}, and \ref{fig:SphereShells}. In region
    labeled (H$^\star$), in addition to a symmetric shell, several
    similarly sized non-symmetric shells assembled. The white areas
    correspond to regions in which different types of shells without
    any specific symmetry are grown. The unlabeled dark shaded region
    in the upper left corner of the plot with large FvK number and
    large spontaneous radius of curvature corresponds to a region in
    which no pentamers can be formed and the elastic sheet curves
    around to form a cylinder.}
\end{figure}

\begin{figure}
  \includegraphics[width=8.7cm]{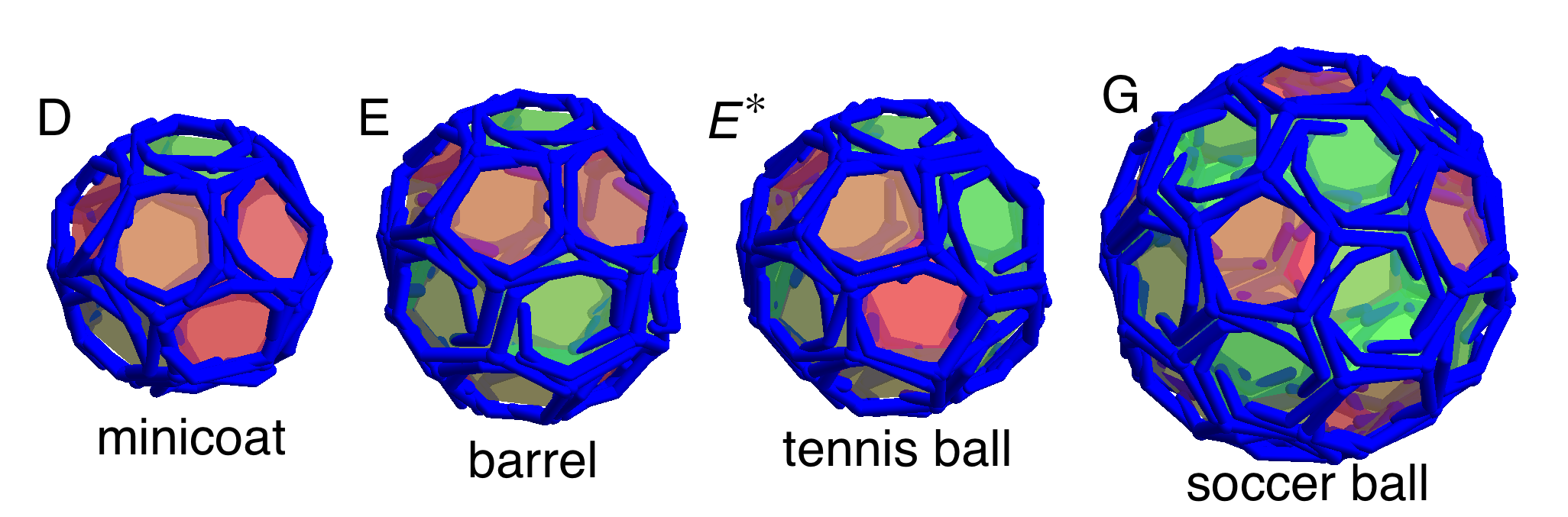}
  \caption{\label{fig:ClathrinShells} (color online) Structures
   presented with the three legged triskelion subunits. These shells
   are very similar to clathrin coats. Of particular interest are the
   symmetric shells seen in {\it in vitro} studies of empty clathrin
   coats, such as the minicoat with 28 subunits and tetrahedral
   symmetry (D), the hexagonal barrel with 36 subunits and $D_6$
   dihedral symmetry (E), the a tennis ball with 36 subunits and
   $D_2$ symmetry (E$^\star$), and a soccer ball with 60 subunits and
   icosahedral symmetry  (G). The (D), (E), (E$^\star$), and ($G$)
   structures correspond to the labeled regions in
   Figs.~\ref{fig:paramspace} and \ref{fig:paramspace2}. The open
   faces are shaded red if they have five sides or green in case of
   six sides. Note that (E$^\star$) only appears in stochastic simulations (see the text). }
\end{figure}

\begin{figure}
  \includegraphics[width=8.7cm]{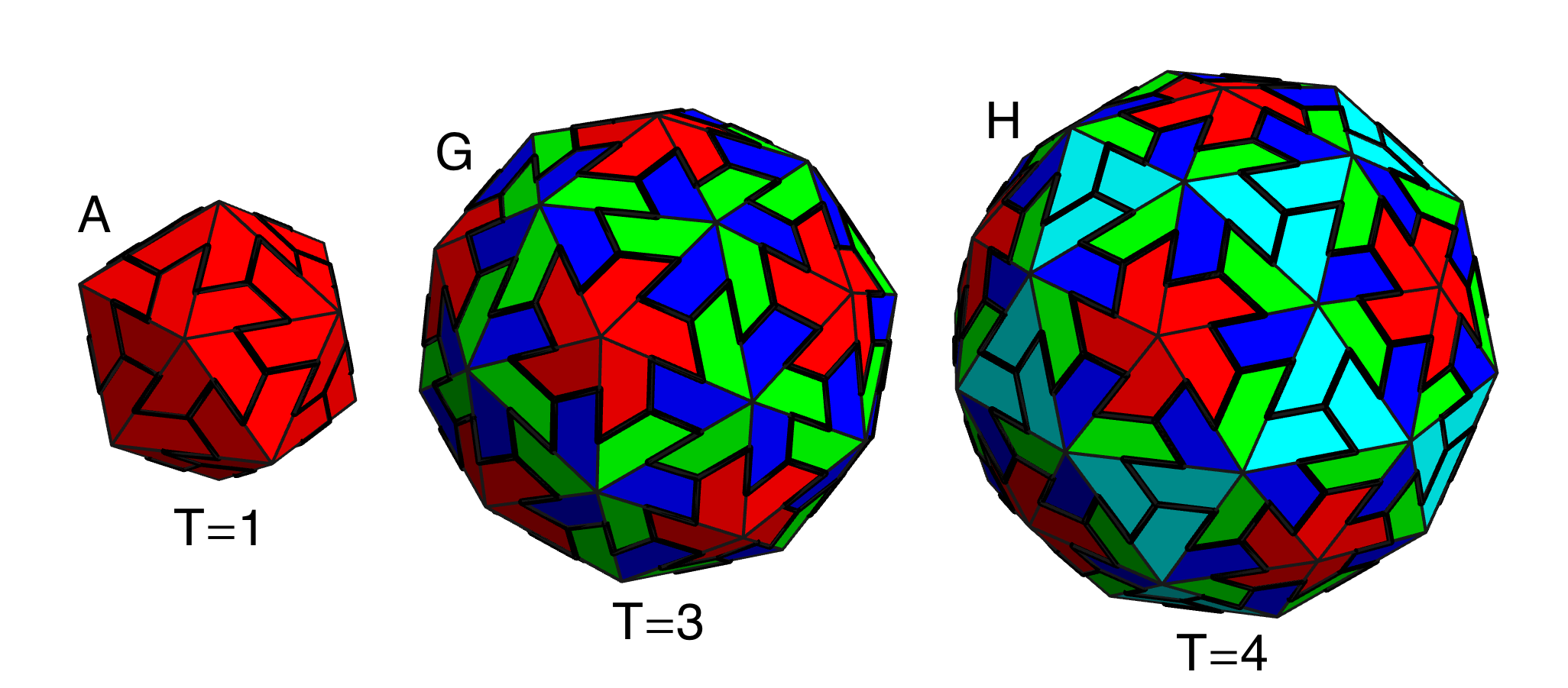}
  \caption{\label{fig:VirusShells} (color online) There are three
    shells that are grown with icosahedral symmetry, reminiscent of
    viral shells. These shells contain (A) 20, (G) 60, and (H) 80
    subunits corresponding to T=1, T=3, and T=4 viral shells based on
    the Caspar-Klug classification of icosahedral shells. Larger
    icosahedral shells are not seen in simulated non-reversible
    growth model, perhaps explaining why scaffolding is needed for
    all larger icosahedral viruses.}
\end{figure}

\begin{figure}
  \includegraphics[width=8.7cm]{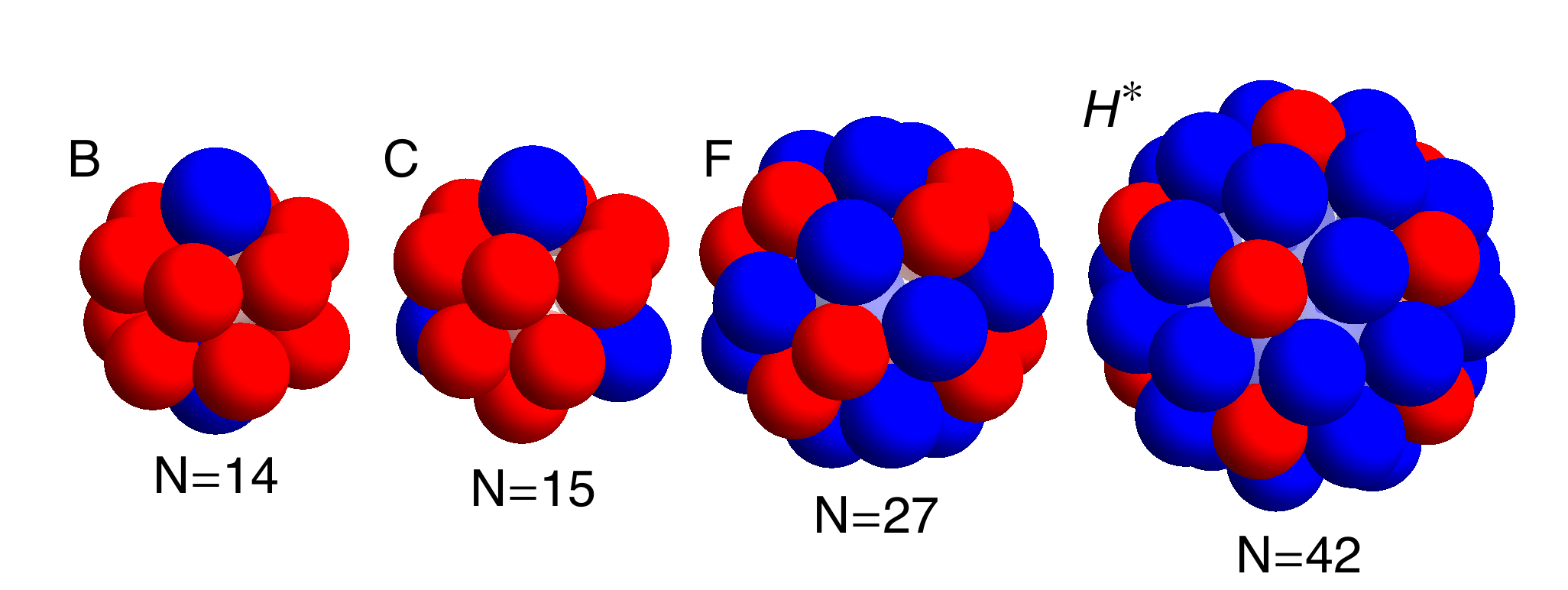}
  \caption{\label{fig:SphereShells} Additional shells that match the
    results of simulations presented in Chen {\it et al.}
    \cite{Chen:2007a, Chen:2007b}. These include (b) N=14 with
    $D_{5h}$ symmetry, (c) N=15 with $D_{3h}$, (g) N=27 with $S_6$
    symmetry, and ($h^\star$) N=42 with $D_{5h}$ symmetry.}
\end{figure}

We investigated the growth of shells as described in the previous
section for several values of dimensionless parameters
$0.1<\bar{\gamma}<10$ and $1.1 < R_0/b_0 < 1.8$. As illustrated in
Fig.~\ref{fig:paramspace}, we found that over a large range of
parameter space, only a few different structures formed.  Most of
these structures had high symmetry and were robust, insensitive to
small changes in the subunit's spontaneous radius of curvature or
mechanical properties. The shaded regions labeled (A) through (H)
in Fig.~\ref{fig:paramspace} correspond to the symmetric shells
presented in Figs.~\ref{fig:ClathrinShells}, \ref{fig:VirusShells}
and \ref{fig:SphereShells}, plotted with the subunits shown in
Figs.~\ref{fig:ShellTypes}b, \ref{fig:ShellTypes}a, and
\ref{fig:ShellTypes}c, respectively. The reason we plotted the
shells with different types of subunits will become clear below.  All
symmetric shells are shown together with the same type of subunit in
Fig.~\ref{fig:SymmetricShells} in the appendix. The number of
subunits $n_s$ and vertices $N$ for each structure are given in the
caption of Fig.~\ref{fig:SymmetricShells}.

To investigate the effect on the shell size of FvK number and the
spontaneous radius of curvature we first studied in detail the growth
of small shells for a fixed value of the FvK number and a range of
spontaneous radius of curvatures.  Note that the FvK number is the
ratio of the stretching to bending modulus, and as such a large FvK
number corresponds to a system in which the subunits significantly
resist deformation from the shape of an equilateral triangle, while a
small one corresponds to a system in which the subunits can be easily
deformed.

\begin{figure}
  \includegraphics{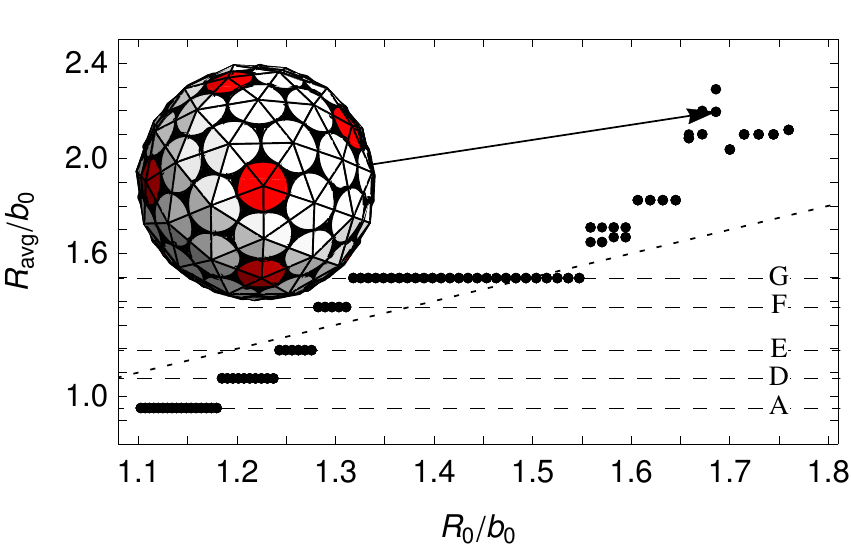}
  \caption{\label{fig:RadiusvsRadiusPlot2} A plot of the average
    radius $R_{\text{avg}}/b_0$, defined as the radius of gyration of
    the vertices of the shell, versus the spontaneous radius of
    curvature $R_0/b_0$. The FvK number was fixed at a smaller value
    of $\bar{\gamma}=2$, which corresponds to a mildly flexible
    subunit. The average radius of the symmetric shells (A), (D), (E)
    (F) and (G) are shown with the horizontal dashed lines. It is
    possible to see a stair step like feature with each flat step
    corresponding to a single type of symmetric shell. With the
    exception of the shells (A) and (F), the corresponding shell
    types are all seen in studies of empty clathrin coats performed
    {\it in vitro}. Representative shells are shown in
    Fig.~\ref{fig:ClathrinShells}.  Note that the small radius of
    curvature of the shell (A) and the unreliable formation of shell
    (F) in the stochastic simulations could explain the absence of
    these shells in the experiments.}
    \end{figure}

The results of our simulations for an intermediate value of FvK
number $\gamma=2$ are plotted in Fig.~\ref{fig:RadiusvsRadiusPlot2},
which shows the average radius of completed shells vs. the
spontaneous radius of curvature. The stair-step like feature in
Fig.~\ref{fig:RadiusvsRadiusPlot2} shows that we found a discrete set
of shells, formed according to the spontaneous radius of curvature.
In general for FvK numbers $\gamma > 1.5$, as the spontaneous radius
of curvature increased, we observed discrete jumps from one symmetric
shell type to another. 

Figure~\ref{fig:ClathrinShells} illustrates some of the structures
associated with the labelled flat steps in
Fig.~\ref{fig:RadiusvsRadiusPlot2} for $\gamma=2$. These shells are
similar to those observed in clathrin vesicles: the minicoat, the
hexagonal barrel, the tennis ball, and the soccer ball shells
described in Fig.~\ref{fig:ClathrinShells}.  The subunit of clathrin
shells is a triskelion molecule, shown in the upper left corner of
Fig.~\ref{fig:ShellTypes}b, whose legs bond anti-parallel. While the
molecules are relatively stiff, the long legs as well as the
variability of the bond angle between the legs can lead to an
intermediate FvK number $\gamma=2$, and as a consequence a range of
shells as seen in Fig.~\ref{fig:ClathrinShells} are grown.
It is believed that the radius of curvature of the clathrin shells is
determined by the cargo\cite{Fotin:2004a} explaining further the
variability of the bond angle and the small FvK number. For empty
clathrin vesicles, it should be the preferred spontaneous curvature between
clathrin subunits that defines the size of the smallest clathrin
shells observed in the experiments, the so-called minicoat vesicle
with 28 subunits, and not a dodecahedron with 20 subunits.

Figure \ref{fig:RadiusvsRadiusPlot2} also shows that for FvK number
$\gamma=2$, as the spontaneous radius of curvature increased beyond
$R_0/b_0 > 1.55$, irregular shells without any underlying symmetry
were assembled. An example of such an irregular shell is
shown in the inset of Fig.~\ref{fig:RadiusvsRadiusPlot2}. The
irregular shells in this regime are more spherical in comparison to
those with a larger FvK number as discussed below, and average radius
of the completed shells closely matches the spontaneous radius of
curvature of the subunits.

In contrast to clathrin vesicles and the shells obtained in
theoretical simulations constructed from cones, spheres, or disk
\cite{Zandi:2004a,Chen:2007a, Chen:2007b}, which assemble into
shells with different symmetries, viruses seem to only form
structures with icosahedral symmetry {\it in vivo}. To explain this
feature, we varied the FvK number and examined in detail its role on
the shell symmetry. Our findings show that the FvK number can indeed
explain the difference between shells appearing in viral capsids vs.
clathrin coats.

\begin{figure}
  \includegraphics{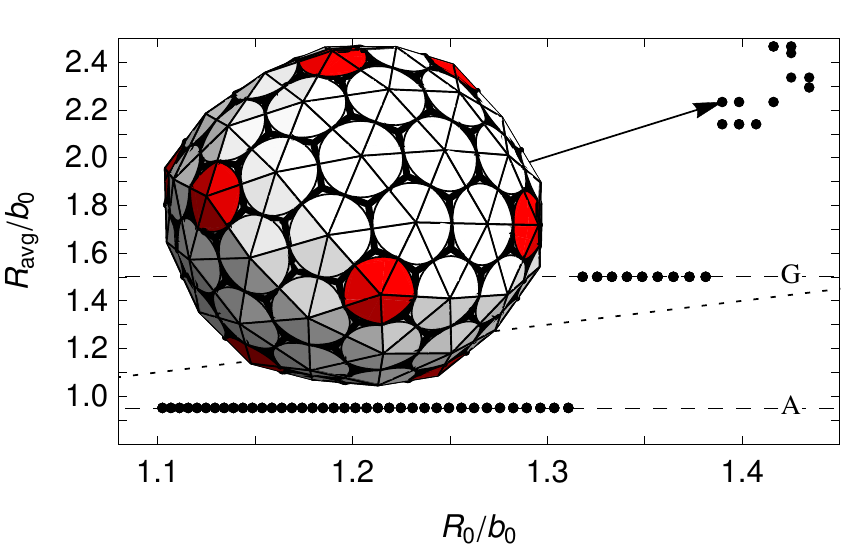}
  \caption{\label{fig:RadiusvsRadiusPlot1} A plot of the average
    radius $R_{\text{avg}}/b_0$ versus the spontaneous radius of
    curvature $R_0/b_0$. The FvK number is fixed at a large value of
    $\bar{\gamma}$=10 corresponding to a very stiff subunit
    (difficult to deform). The two large flat regions seen in the
    plot correspond to small icosahedral structures shell types (A)
    (G) in Fig.~\ref{fig:VirusShells}. For spontaneous radius of
    curvatures larger than $R_0/b_0 > 1.38$, large non-symmetric and
    non-spherical shells are grown. An example is shown in
    the inset.}
\end{figure}

For a fixed large FvK number $\gamma=10$, the size of shells grown as
a function of the spontaneous radius of curvature is illustrated in
Fig.~\ref{fig:RadiusvsRadiusPlot1}. We found two large flat regions,
where rather considerable changes in spontaneous radius of curvature
have no effect at all on the size of shell. These two flat steps
correspond to the shells A and G with icosahedral symmetry shown in
Fig.~\ref{fig:VirusShells}. If the subunits are considered as capsid
protein trimers these correspond to T=1 and T=3 icosahedral shells
based on the Caspar-Klug classification, see
Fig.~\ref{fig:VirusShells}. As the spontaneous radius of curvature
was increased $R_0/b > 1.38$, we found that large irregular oblong
and ellipsoidal shells formed as illustrated in the inset to
Fig.~\ref{fig:RadiusvsRadiusPlot1}. As the spontaneous radius of
curvature was increased even further the formation of pentamers
(which are required in order to from closed shells) was suppressed
completely, and only rolled flat sheets were grown.

\begin{figure}
  \includegraphics{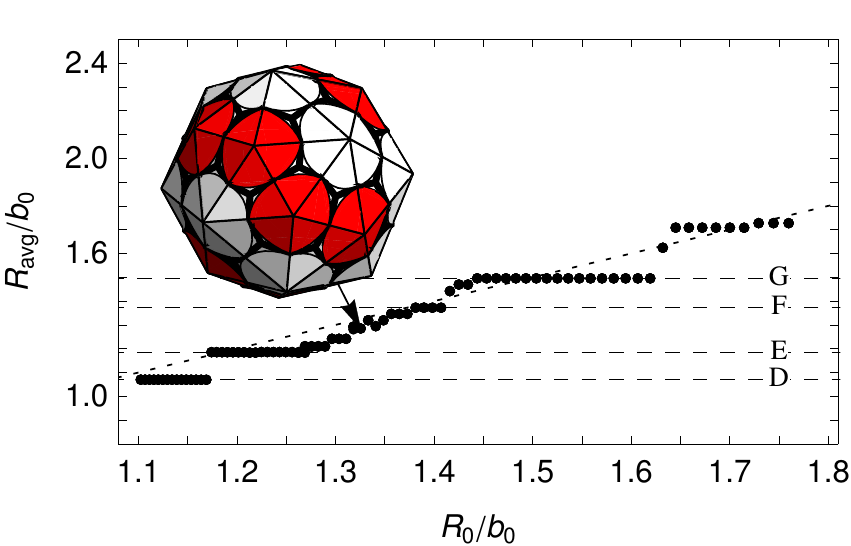}
  \caption{\label{fig:RadiusvsRadiusPlot3} A plot of the average
    radius $R_{\text{avg}}/b_0$ vs. the spontaneous radius of
    curvature $R_0/b_0$. The FvK number is fixed at a small value of
    $\bar{\gamma}$=0.25 corresponding to a flexible subunit. The
    average radius of the symmetric shells (D), (E) (F) and (G) (see Fig.~\ref{fig:SymmetricShells}) are
    shown with the horizontal dashed lines. The dotted
    line corresponds to the line where $R_{\text{avg}}=R_0$. Note
    that between plateaus corresponding to shell types (E), (F) and (G) there are several
    non-symmetric shells whose average radius is closer to the
    spontaneous radius. An example of non-symmetric shell is shown in the
    inset.}
\end{figure}

Comparison of Figs.~\ref{fig:RadiusvsRadiusPlot2} and
\ref{fig:RadiusvsRadiusPlot1} shows that the more flexible subunits
(lower FvK number) allowed for a larger variety of shells to grow. In
fact, Fig.~\ref{fig:paramspace} shows that for $\gamma=0.5$ a few
more structures in addition to those associated with clathrin
vesicles ($\gamma=2$) or viral shells ($\gamma=10$) were grown. These
additional structures are plotted in Fig.~\ref{fig:SphereShells}.
Interestingly these shells were also observed in the molecular
dynamic simulation of conical particles \cite{Chen:2007a} or
spherical particles with convexity constraints \cite{Chen:2007b} as
explained in the introduction. As the FvK number was decreased
$\gamma < 1.5$,  a new behavior emerged: The sharp transition between
shell types (E) and (F) disappeared, and as the FvK number was
further decreased below $\gamma < 0.4$ the sharp transition between
shell types (F) and (G) also disappeared, see
Fig.~\ref{fig:RadiusvsRadiusPlot3}. In between these regions small
relatively spherical non-symmetric shells whose radii closely match
the spontaneous radius of curvature were grown.

Our results illustrated in Figs.~\ref{fig:paramspace},
\ref{fig:RadiusvsRadiusPlot2}, \ref{fig:RadiusvsRadiusPlot1}, and
\ref{fig:RadiusvsRadiusPlot3} clearly show that icosahedral
structures are robust and appeared for all values of FvK numbers. The
fact that most small spherical viruses only adopt T=1 and T=3 icosahedral structures
and not other symmetric shells illustrated in
Figs.~\ref{fig:ClathrinShells} and \ref{fig:SphereShells} indicate
that viral coat proteins should take on a rigid tertiary structure
and be difficult to deform as it is the case of $\gamma=10$
structures illustrated in Fig.~\ref{fig:RadiusvsRadiusPlot1} .

It is interesting to note that only at the smaller FvK numbers,
$\bar{\gamma}<0.25$, we observed the $n_s=80$ icosahedral shell,
corresponding to a T=4 capsid in the Caspar-Klug classification, see
Fig.~\ref{fig:VirusShells}. We found two other structures with
$n_s=80$ subunits at intermediate FvK numbers, $0.25 < \bar{\gamma} <
2$. However these shells either displayed a 5-fold $D_{5}$ symmetry
or no symmetry at all. The two non-icosahedral $n_s=80$ shells were
not as spherical as the icosahedral one. The $D_{5}$ structure
(Fig.~\ref{fig:SphereShells}H$^\star$), corresponding to the $N=42$
magic number shells also seen in Refs.~\cite{Chen:2007a, Chen:2007b},
was slightly prolate spheroidal, while the completely non-symmetric
structure was ellipsoidal. It is worth mentioning that while
icosahedral structures corresponding to $n_s=20$ and $n_s=60$ covered
large areas of the parameter space in Fig.~\ref{fig:paramspace} and
are still robust in the stochastic growth model (see below), the
$n_s=80$ icosahedral shell only appeared in a small region of the
phase space and was not as robust. This might indicate that forming a
T=4 capsid with triangular subunits similar to those employed in the
simulations done in this paper is difficult. Also this could explain
the relevant abundance of T=1 and T=3 icosahedral virus capsids
compared to the T=4 ones observed in nature. Note that $N$=42
icosahedral structures did not appear in the simulations of
Refs.~\cite{Chen:2007a, Chen:2007b} (instead they found the $D_5$
shell mentioned above) and appeared in Ref.~\cite{Zandi:2004a} only
if the overall system was under pressure.

\begin{figure}
  \begin{center}
    \includegraphics{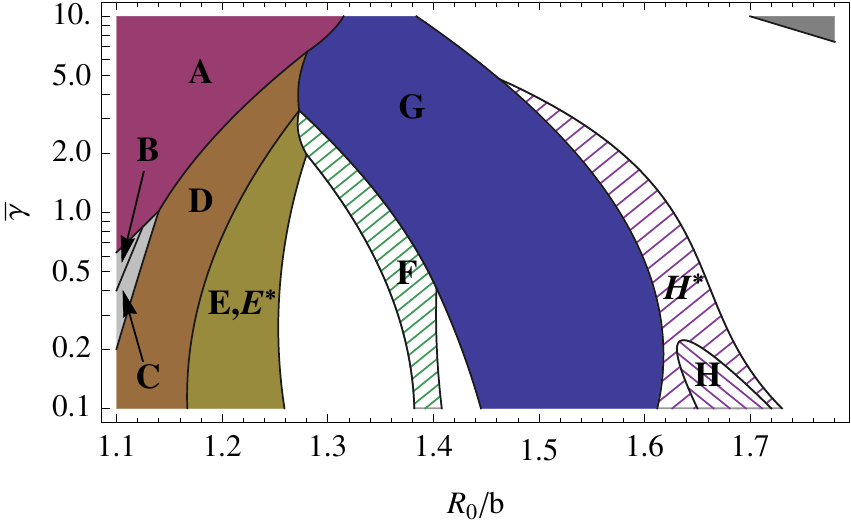}
  \end{center}
  \caption{\label{fig:paramspace2} (color online) Phase diagram for
    the stochastic model with $\sigma = 20^\circ$ showing the
    different kind of shells assembled for values of the
    dimensionless parameters $\bar{\gamma}$ and $R_0/b_0$. In spite
    of the stochastic growth, only a single shell type is grown in
    each solid shaded region. However, in the hashed regions several
    similarly sized non-symmetric shells are formed in addition to
    the symmetric shell. The white areas correspond to regions
    where different types of shells without any regular symmetry
    are assembled. The unlabeled dark shaded area in the upper left
    corner of the plot with a large FvK number and  large spontaneous
    radius of curvature corresponds to a region in which no pentamers
    can be formed and the elastic sheet curves around to form a
    cylinder.}
\end{figure}

In addition to the deterministic simulations, we studied the growth
of shells for the stochastic model with the width of the distribution
$\sigma=20^{\circ}$ (see Eq.~\ref{eq:stochastic}), which was the
smallest value of $\sigma$ for which there was an appreciable
difference between the stochastic and deterministic models. We found
that the symmetric shells still reliably grew in the stochastic model
exactly in the same regions as they appeared in the deterministic
one, as shown in Fig.~\ref{fig:paramspace2}. This indicates that the
stochastic choice of where to add the next subunit did not have a
strong effect on the final structure of the symmetric shells. This
was not, however, the case for the irregular shells that correspond
to the white areas of Fig.~\ref{fig:paramspace}. In these regions of
parameter space multiple and new irregular shells were formed
compared to the structures assembled in the deterministic model.

There were a few exceptions to the robustness of the assembly of
symmetric shells in the stochastic model. Both the $N=27$
(Fig.~\ref{fig:SphereShells}F) and T=4 (Fig.~\ref{fig:VirusShells}H)
shells were not reliably formed in the stochastic growth model in
that, in addition to the symmetric structures, similar sized non-
symmetric shells were formed in the same region of parameter space.
Further, one additional symmetric shell was observed in the
stochastic model that did not appear in the deterministic one; the
tennis ball shell (Fig.~\ref{fig:ClathrinShells}E$^\star$). The
barrel (Fig.~\ref{fig:ClathrinShells}E) and tennis ball both have the
same number of clathrin subunits, $n_s=36$, which consist of 12
pentagonal and 8 hexagonal faces. Shell (E), which was the only shell
seen with $n_s=36$ in the deterministic simulation, contains two
rings of six pentagons separated by a ring of 6 hexagons. Shell
(E$^\star$) has the 12 pentagons all connected in a line that winds
around the shell in the same manner as the seam on a tennis ball, and
the 8 hexagons lay in two patches of 4. Note that both structures were
observed in the experimental studies of assembly of clathrin
coats\cite{Crowther:1976a}.

Remarkably, despite the non-equilibrium and stochastic assembly
pathways, there were vast regions in the parameter space displaying
shells with a high degree of symmetry, see Fig.~\ref{fig:paramspace}
and Fig.~\ref{fig:paramspace2}.  This is due to the interplay between
the spontaneous radius of curvature and the response of the elastic
triangular network to local stresses. The radius of curvature of a
regular icosahedron made up of 20 triangular subunits is $R_0/b_0
\approx 0.915$ while sheets made of only hexamers are perfectly flat
with $R_0/b_0 = \infty$. For the range of $R_0$ studied ($1.1 <
R_0/b_0 < 1.8$) in this paper, the introduction of a pentamer creates
too much local curvature while the addition of a hexamer will induce
too little curvature. Depending which type of capsomers creates more
stress we found that the local distortion tended to discourage the
formation of similar capsomers in the immediate vicinity of each
other.  It is this effective repulsion that gives rise to the
symmetric structures observed in different systems.  Small
spontaneous radius of curvature ($R_0/b_0<1.3$) promotes formation of
pentamers and repulsion between hexamers, leading to the small shells
with tetrahedral and dihedral symmetries. Larger $R_0/b_0$, on the
other hand, favors formation of hexamers and repulsion between
pentamers resulting in, for example the $n_s=60$ and $n_s=80$ subunit
shells with icosahedral symmetry.

For large values of spontaneous radius of curvature we were indeed
able to obtain similar results to the work of Hicks and
Henley\cite{Hicks:2006a} and Levandovsky and
Zandi\cite{Levandovsky:2010a, Yu:2013a} both of which used slightly
different variants of the model studied in this work. For larger
spontaneous radius of curvatures, we obtained mostly irregular capsids
with many of the larger capsids displaying defects identical to those
observed in Ref.~\cite{Hicks:2006a}. For large shells pertinent to
retroviral shells and a fixed FvK number, we found a similar behavior
to that seen in Ref.~\cite{Levandovsky:2010a}, {\it i.e.}, as
spontaneous radius of curvature increased, the type of capsids formed
changes from the irregular spheroidal capsids to larger irregular
capsids including cone shaped structures to finally tube shaped
capsids.


\section*{CONCLUDING REMARKS}

In summary we used a minimal non-equilibrium model to study
the growth of nano-structures constructed with identical subunits. We
found that large spontaneous curvatures between subunits leads to the
growth of many small highly symmetric shells, consistent with those
observed in nature. The formation of most of the symmetric shells was
robust; they were formed over large areas of the parameter space
despite the non-reversible and stochastic growth pathways used.

While the mechanical properties of the subunits, described by the
single dimensionless FvK number, determined which set of shells was
allowed to form as shown in Figs.~\ref{fig:ClathrinShells},
\ref{fig:VirusShells}, and \ref{fig:SphereShells}, the spontaneous
radius of curvature determined the size of the shells. For fixed FvK
number, the impact of spontaneous curvature on the size and structure
of shells was demonstrated in Figs.~\ref{fig:RadiusvsRadiusPlot2},
\ref{fig:RadiusvsRadiusPlot1}, and \ref{fig:RadiusvsRadiusPlot3}.
These plots clearly indicate that the FvK number $\bar{\gamma}$ can
explain why shells formed from conical particles\cite{Chen:2007a} or
clathrin shells\cite{Crowther:1976a} display many more magic numbers
than the shells formed from viral capsid proteins.

While a large $\bar{\gamma}$ implies that the subunits are difficult
to deform from the preferred shape of an equilateral triangle,  small
$\bar{\gamma}$ implies more flexible subunits. The conical particles
used in Ref.~\cite{Chen:2007a} do not have any explicit triangular
symmetry, the underlying hexagonal symmetry only appears due to the
close packing of circles in a plane, See Fig.~\ref{fig:ShellTypes}.
The lack of explicit triangular symmetry leads to a smaller FvK
number $\bar{\gamma}$, which allows for more diverse shells to be
formed. The long legs and variability of the bonding angle of the
triskelion molecules that make up the clathrin results into a more
flexible molecule and thus a small FvK number compared to that of
viral shells. In contrast to the conical particles employed in
simulations and the triskelion molecules in clathrin shells, we
expect the viral capsid proteins assume a large FvK number as they
only form structures with icosahedral symmetry. It is worth
mentioning that while it is wildly accepted that the shells with
icosahedral symmetry have the lowest free energy
structures\cite{Zandi:2004a,Chen:2007a}, our work shows that small
highly symmetric shells can be reliably formed under non equilibrium
conditions and over a wide range of parameter space the {\it local}
minimum free energy path leads to a {\it global} minimum free energy
structure.

A careful analysis of all the structures formed shows that in the
range of curvatures studied in this paper, the mismatch between the
desired spontaneous radius of curvature and the local radius of
curvature induced by a pentamer or hexamer tends to suppress the
formation of similar types of capsomers in the vicinity of each
other. This suppression tends to lead to the small shells being
highly symmetric in nature. This effect seems to be short ranged,
which can explain both why larger clathrin vesicles display no
symmetry and why larger viruses need scaffolding proteins or other
mechanisms to form shells with icosahedral symmetry. While FvK
equations can be solved in the absence of spontaneous curvature and
give an estimate for the repulsion between
pentamers\cite{Lidmar:03a}, in the presence of spontaneous curvature,
the highly non-linear nature of equations makes solving these
equations completely non-trivial.

Understanding the mechanisms and assembly pathways of formation of
highly symmetric robust shells could have significant impact in the
development of anti-viral therapies and in design of novel biomimetic
materials.

\section*{APPENDIX A}
\renewcommand{\thefigure}{A.\arabic{figure}}
\setcounter{figure}{0}    
The symmetric shells assembled through the non-reversible addition of
triangular subunits in our work were found to match the symmetric
structures in three disparate systems: viral capsids, clathrin coats,
and theoretical simulations involving identical cone shaped
particles. The shells displayed in Figs.~\ref{fig:ClathrinShells},
\ref{fig:VirusShells}, and \ref{fig:SphereShells} in the main text
are plotted with the relevant subunits to each case in order to best
show the match between our work and these systems. However, to
provide a consistent comparison between all symmetric shells, we
illustrate them all using the same subunit, see
Fig.~\ref{fig:SymmetricShells}. As clearly revealed in the figure, as
the size of shell increases, the strong repulsion between hexamers
diminishes and repulsion between pentamers becomes stronger, leading
to the shell with different symmetry type. Videos of the step by step
growth of each shell is presented in the SI.

\begin{figure*}
  \includegraphics[width=\textwidth]{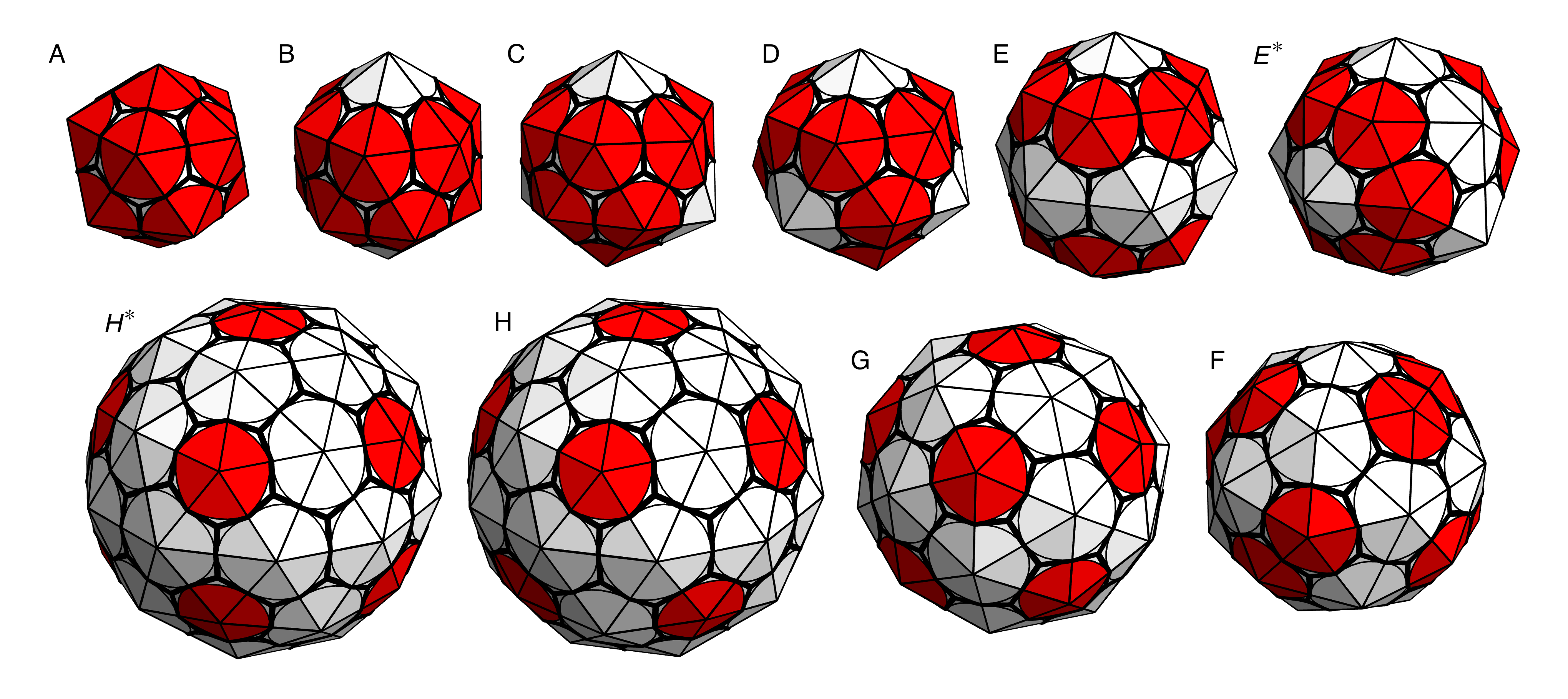}
  \caption{\label{fig:SymmetricShells} (color online) The set of
    small symmetric shells that are formed in the similarly labeled
   shaded regions in the parameter space plots in Figs.~\ref{fig:paramspace} and \ref{fig:paramspace2}. The
   subunits are shown as triangles. The pentamers (points with 5
   triangles attached) appear darker (red online) in the figure. The
   shells (A) through (H) have $n_s$= 20, 24, 26, 28, 36, 50, 60, and
   80 subunits or $N$= 12, 14, 15,1 6,20, 37, 32, and 42 vertices
   (pentamers and hexamers) respectively. It should be noted that the
   sets of shells (E) and ($E^\star$), and (H) and ($H^\star$) have
   the same number of subunits but different symmetries. The
   symmetries of the shells from left to right and top to bottom are
   icosahedral, $D_6$, $D_3$, tetrahedral, $D_6$, $D_2$, $D_5$,
   icosahedral, icosahedral, and $S_6$.}
\end{figure*}

\section*{SUPPLEMENTARY MATERIAL}

\ack{An online supplement to this article can be found by visiting BJ
Online at http://www.biophysj.org.}\vspace*{-3pt}

\section*{ACKNOWLEDGMENTS}
The authors would like to thank Gonca Erdemci-Tandogan for many
helpful discussions. This work was supported by the National Science
Foundation through Grant No. DMR-1310687.

\bibliographystyle{biophysj}
\bibliography{shell}

\end{document}